# Magnetic Interactions Between Radical Pairs in Chiral Graphene Nanoribbons

Tao Wang,[1,2,*,†] Sofia Sanz,[1,*] Jesús Castro-Esteban,[3] James Lawrence,[1,2] Alejandro Berdonces-Layunta,[1,2] Mohammed S. G. Mohammed,[1,2] Manuel Vilas-Varela,[3] Martina Corso,[1,2] Diego Peña,[3,‡] Thomas Frederiksen,[1,4,§] and Dimas G. de Oteyza[1,2,4,∥]

[1]*Donostia International Physics Center, San Sebastián, Spain*

[2]*Centro de Fisica de Materiales CFM / MPC, CSIC-UPV/EHU, San Sebastián, 20018, Spain*

[3]*Centro Singular de Investigación en Química Biolóxica e Materiais Moleculares (CiQUS) and Departamento de Química Orgánica, Universidade de Santiago de Compostela, Santiago de Compostela, Spain*

[4]*Ikerbasque, Basque Foundation for Science, Bilbao, Spain*

**Abstract:** Magnetic graphene nanoribbons (GNRs) have become promising candidates for future applications, including quantum technologies. Here, we characterize magnetic states hosted by chiral graphene nanoribbons (chGNRs). The substitution of a hydrogen atom at the chGNR edge by a ketone group effectively adds one $p_z$ electron to the π-electron network, thus producing an unpaired π radical. A closely related scenario occurs for regular ketone-functionalized chGNRs in which one oxygen atom is missing. Two such radical states can interact *via* exchange coupling and we study those interactions as a function of their relative position, which includes a remarkable dependence on the chirality, as well as on the nature of the surrounding GNR, *i.e.*, with or without ketone functionalization. In addition, we determine the parameters whereby this type of systems with oxygen heteroatoms can be adequately described within the widely used mean-field Hubbard model. Altogether, we provide new insights to both theoretically model and devise GNR-based nanostructures with tunable magnetic properties.

Magnetic carbon nanostructures exhibit attractive properties that differentiate them from the conventional magnetic systems relying on *d* or *f* states, as are the weaker spin-orbit coupling and a larger spin delocalization [1–5]. The open-shell character and the corresponding magnetic properties may appear in graphene nanoflakes (GNFs) and nanoribbons (GNRs) with certain topologies for a number of reasons. Firstly, as predicted by Lieb's theorem [6], sublattice imbalance in a bipartite lattice leads to a net spin inside the nanographene, as occurs *e.g.* in triangulene [7]. Even with balanced sublattices, topological frustration in GNFs may hinder pairing all $p_z$ electrons and result in open-shell structures [8]. In addition, if the Coulomb repulsion between valence electrons is comparable to the band gap between molecular frontier orbitals, it can prompt the system to host singly occupied orbitals [9–13]. Finally, a net spin can be introduced to GNFs and GNRs by simply adding/removing an odd number of π-electrons into/from the system [14,15]. Besides charge transfer scenarios [16–18], this can also be achieved by the insertion of odd-membered rings [15–18], by an $sp^2$ to $sp^3$ rehybridization as driven *e.g.* by hydrogenation, or by heteroatom substitutions. The former two approaches have been increasingly employed [10,15–19] but only few examples elucidate the latter one [20] though with many theoretical predictions [14,21,22].

The synthesis and characterization of graphene π-magnetism has seen great advances thanks to the recent development of on-surface synthesis [10,16,17,20,23–37]. The topologies of GNFs and GNRs can be precisely tuned by the rational design of precursor molecules. Because of their delocalized π-magnetism, magnetic GNFs and GNRs are ideal candidates for the construction of interacting electron spin systems. Most of the reported works are focused on the interactions between radicals on GNFs [23,37], whereas the engineering of exchange-coupled spins on extended systems like GNRs, though highly desired, is more scarce [20,25,38].

Herein, we report the generation of net spins on chiral GNRs (chGNRs) by ketone functionalization [Fig. 1(a)]. Replacing a CH at the pristine chGNR edge by a C=O (carbonyl) adds one π-electron to the system. As a result, the odd number of total electrons causes the appearance of a π-radical. The same is applicable for regularly ketone-functionalized chGNRs [39] if a C=O is replaced by a CH. The introduction of a second radical may either maintain the system's open-shell character or result in their hybridization into a closed-shell structure [31,37]. Based on the above, a variety of radical pairs with different geometries were analyzed in our experiments. Combining scanning tunneling microscopy (STM) with theoretical calculations, we reveal how their relative location, the chemical structure of the units surrounding the two radicals and the chirality itself, all influence the spin interactions.

The generation of radical states in chGNRs is schematically displayed in Fig. 1(a) and starts from the readily described synthesis of pristine (P-) and ketone-functionalized (K-) chGNRs on Au(111) [39,40]. In the case of K-chGNRs, a few defective ribbons with one or more missing oxygen atoms normally coexist with intact ribbons [39]. The substitution of an oxygen by a hydrogen atom makes the total number of $p_z$ electrons an odd number, thus producing an unpaired π-radical, as indicated by the blue arrow in Fig. 1(a). In turn, the addition of an oxygen atom to an otherwise P-chGNRs effectively adds one $p_z$ electron and equally brings in a radical. This can be achieved either by exposing P-chGNRs to $O_2$ and post-annealing [41], or removing most of the ketones of K-chGNRs by exposing them to atomic hydrogen and post-annealing [39]. Using these procedures, P- and K-chGNRs with single radicals and with radical pairs were produced. We use the following nomenclature to refer to different radical pairs. Taking the radical marked with the blue arrow in Fig. 1(b) as reference, the other labels denote the position of the second radical [Fig. 1(b)]. S/O represents the two radicals of a pair located at the same/opposite sides of a chGNR. The following number shows how many precursor units separate a radical pair. A '-mark is used to distinguish the differing configurations resulting from the ribbon's chirality. Finally, P- denotes P-chGNRs in which the radicals are caused by additional ketones and K- denotes K-chGNRs with radicals associated to missing ketones. An example P-O1 radical pair is shown in Fig. 1(a).

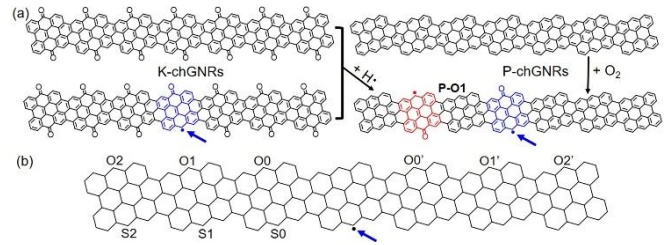

**FIG. 1**. (a) Schematic drawing for the fabrication of magnetic chGNRs. A π-radical is introduced into chGNRs by the addition/removal of an oxygen into/from P-/K-chGNRs. (b) Labels for the radical pairs in chGNRs. Taking the radical marked with a blue arrow as the reference, each pair is labeled depending on the position of the second π-radical.

Figure 2(a) presents the bond-resolving (BR) STM image of a K-chGNR with a single defect using a CO-functionalized probe [42], along with the associated chemical structure. At the low bias value used for BR-STM, the unit cell with a missing ketone exhibits brighter contrast, implying the existence of low-energy states. Figure 2(b) shows the differential conductance spectrum (dI/dV) taken at the marked position in Fig. 2(a). Apart from the highest occupied and lowest unoccupied molecular orbitals (HOMO and LUMO) of a K-chGNR at -1.45 and 0.65 V, associated with the reported valence and conduction band onsets [39], two in-gap states are clearly visible at -0.47 and 0.25 V. The same local density of states (LDOS) distribution at these two energies in dI/dV maps supports their common origin from the singly occupied/unoccupied molecular orbitals (SOMO/SUMO), separated by a 0.72 eV Coulomb gap [36]. In line with the experiments, the spin density calculated for this structure with density functional theory (DFT) [Fig. 2(e)] distributes mostly over the "defective" unit cell, but extends slightly to the adjacent ones with a clear chirality-driven asymmetry [43]. Evidence of this system's net spin $S=1/2$ is detected as a zero-bias peak in the low-energy dI/dV spectra [Fig. 2(f)]. It broadens anomalously fast with temperature and can thus be attributed to a Kondo resonance [23]. The HWHM of the temperature dependent spectra, as extracted from fits to a Frota function [47] [Fig. 2(f)], are further fitted to the Fermi-liquid model [48] [Fig. 2(g)] and result in a Kondo temperature of 66 K. The identical spin distribution of the Kondo map obtained at 5 mV [Fig. 2(g) inset] and of the SOMO/SUMO [Fig. 2(c-d)] further corroborates the origin of the spin density in the SOMO.

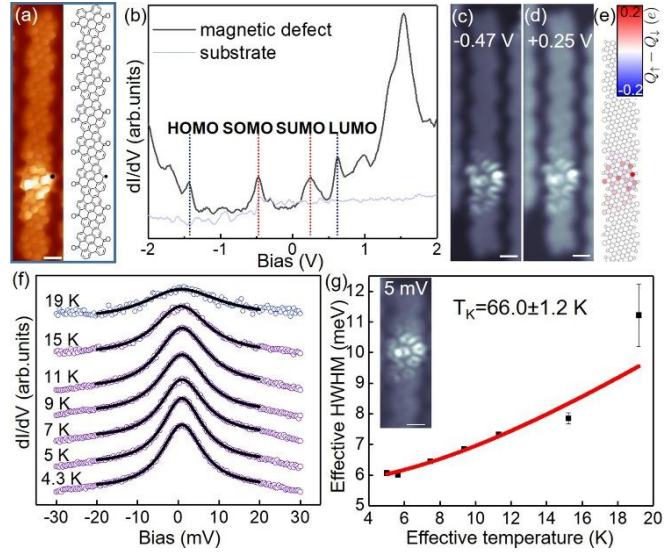

**FIG. 2**. (a) Constant-height BR-STM image of a K-chGNR with single defect (V=5 mV), together with the corresponding chemical structure. (b) dI/dV spectra at the marked position in (a, black line) and the bare Au(111) substrate (grey line), $V_{rms}$=20 mV. (c,d) Constant-height dI/dV maps recorded at V=-0.47 and 0.25 V, respectively. $V_{rms}$=20 mV. (e) DFT calculated spin density distribution in K-chGNR with single defect. (f) Temperature dependence of the Kondo resonance. All the spectra are fitted by a Frota function. (g) Extracted half-width at half-maximum (HWHM) of the Kondo resonances as a function of temperature, fitted by the Fermi-liquid model. The inset shows a Kondo map measured at 5 mV. Scale bars: 0.5 nm.

An equivalent characterization of a single magnetic defect in P-chGNR is presented in the Supplemental Material (SM [43]). Comparing to the K-chGNR case, the SOMO and the associated spin density in P-chGNR show a much more limited contribution on the ketone and mostly distribute over the pristine side of the defective unit cell. The smaller bandgap of P-chGNR as compared to K-chGNR (~0.7 *vs.* ~2.1 eV) [39,49], causes a stronger hybridization of the radical states with the frontier molecular orbitals, resulting in a notably more extended distribution into neighboring unit cells (though with a similar chirality-driven asymmetry) and consequently a lower Coulomb gap of only 0.35 eV.

Next we focus on chGNRs with radical pairs. Whereas a hybridization of the two radicals may result in a closed-shell ground state with a doubly occupied HOMO, the system can also remain open-shell, with the associated SOMOs, if the hybridization energy is lower than the Coulomb repulsion between the corresponding electrons [31,37]. Our DFT calculations (see SM [43]) predict two radicals in closest proximity but opposite GNR sides (*i.e.*, the O0 and O0' cases in both types of chGNRs) to hybridize into a closed-shell ground state, while all the other radical combinations show open-shell ground states. The experimentally observed P-O0, P-O0', and K-O0' cases confirm their closed-shell character and are displayed in the SM [43], with HOMOs and LUMOs exhibiting distinct LDOS distributions, opposed to that expected from SOMO and SUMO.

For the open-shell structures, the exchange interaction between a radical pair can be experimentally accessed from inelastic spin-flip excitations in scanning tunneling spectroscopy [23]. The low-energy dI/dV spectra of radical pairs with a S=1 triplet ground state may exhibit three features: an underscreened Kondo resonance at Fermi level and/or two side steps symmetrically positioned around the Fermi level associated with inelastic triplet-to-singlet spin-flip excitations [16,34]. In contrast, the dI/dV spectra of radical pairs with a S=0 singlet ground state exhibit only the two singlet-to-triplet side steps [16,17,36,37]. As the exchange interaction between the radicals becomes sufficiently weak, equaling singlet and triplet energies, the radicals respond independently from one another and may display only a Kondo resonance. The geometries of the experimentally addressed radical pairs are determined from the BR-STM images displayed in Fig. 3(a,d), along with their spin densities as calculated with DFT. The associated low-energy dI/dV spectra are shown in Fig. 3(b,c). In agreement with the above, further confirmed with fits to the spectra (see SM [43]) with a perturbative approach up to third order for two coupled S=1/2 systems using the code from M. Ternes [50], we conclude the exchange coupling energies and relative alignments for each radical pair as summarized in Table I. Table I also includes the values obtained from theoretical calculations. The good match of the DFT results with the experiments underlines the predictive power of the calculations on radical pair geometries not accessed experimentally.

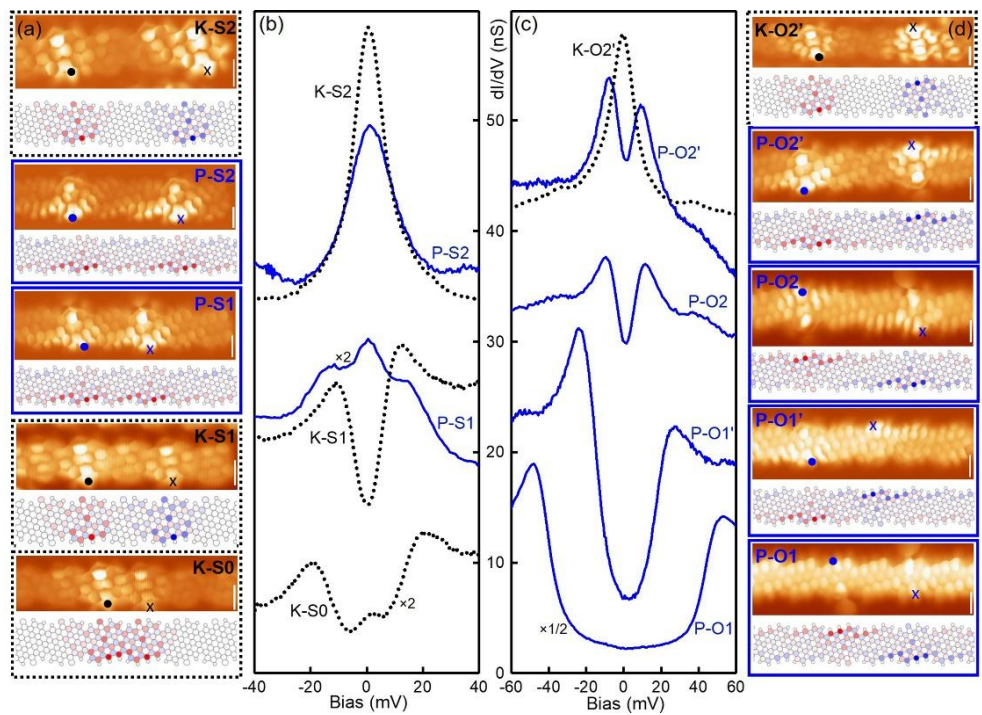

**FIG. 3**. (a,d) BR-STM images (V=5 mV) of experimentally obtained radical pairs at same (a) and opposite sides (d) on P- and K-chGNRs, together with DFT calculated spin-polarized LDOS for each case (ground states). (b,c) dI/dV spectra measured at the marked positions in STM images (see comparative spectra at the position of the second radical marked with an x in the SM [43]). The blue solid and black dotted curves refer to radical pairs on P- and K-chGNRs, respectively. P-O1 shows the original signal and other spectra were vertically shifted to match the positions of STM images. Lock-in amplitude: 2 mV for P-O1, 1.5 mV for K-S0, 1 mV for other cases. Scale bars: 0.5 nm.

**Table I**. Exchange-interaction strength J (in meV) obtained from experiments and theoretical calculations for multiple radical pairs. CS denotes closed-shell structure. Positive (negative) J values indicate preferred AFM (FM) alignment.

| Type | P-S0 | P-S1 | P-S2 | P-O0 | P-O0' | P-O1 | P-O1' | P-O2 | P-O2' | K-S0 | K-S1 | K-S2 | K-O0 | K-O0' | K-O1 | K-O1' | K-O2 | K-O2' |
|---|---|---|---|---|---|---|---|---|---|---|---|---|---|---|---|---|---|---|
| Experiment | | -13.8 | -2.3 | CS | CS | 42.7 | 17.5 | 5.8 | 4.7 | -13.3 | 6.2 | 0 | | | | | | 0 |
| DFT | -31.4 | -18.4 | -11.0 | CS | CS | 35.8 | 28.0 | 11.3 | 11.8 | -24 | 6.4 | 0.4 | CS | CS | 29.1 | -0.5 | 0.1 | 0.7 |
| MFH-TB(1) | -44.0 | -15.5 | -7.2 | CS | 108.0 | 29.0 | 21.4 | 7.8 | 7.8 | -50.9 | -6.6 | -0.5 | CS | 109.1 | 14.4 | 2.7 | 0.9 | 0.3 |
| MFH-TB(2) | -23.2 | -12.8 | -6.4 | CS | CS | 28.8 | 20.4 | 6.9 | 7.1 | -19.0 | 2.1 | 0.1 | CS | CS | 20.9 | 0 | 0.5 | 0 |

Focusing on the exchange coupling J, it scales inversely with the spatial separation for both P- and K-chGNRs cases. That is, radical pairs separated by less unit cells show larger J values due to the greater overlap of the radical states' wavefunctions and the associated spin density [Fig. 3(a,d)]. Interestingly, J shows a remarkable dependence on the chirality, as observed experimentally with P-O1 and P-O1' and predicted theoretically also for K-O1 and K-O1' (Table I). The chirality-driven asymmetric extension of the SOMO wavefunction (and of the associated spin density) into neighboring unit-cells strongly affects their overlap. For example, whereas the spin density of radical states in P-O1 mostly extends toward the central unit between the radical pairs, it dominantly extends away from each other in P-O1' [Fig. 3(d)]. As a consequence, the former shows a much stronger exchange coupling than the latter (42.7 *vs*. 17.5 meV). The exchange coupling strength also varies substantially with the nature of the surrounding GNR, as exemplified here with the presence or absence of the edge functionalization by ketones. Radical states in K-chGNRs extend much less than those in P-chGNRs, promoting in the latter a larger overlap and thus larger J for the same radical pair geometry (Fig. 3 and Table I).

As for the spin's relative alignment, a generally applicable assumption that relies on the preferred antiferromagnetic alignment of electrons in chemical bonds is that, for alternant graphene nanostructures, each of the two sublattices hosts $p_z$ electrons with spin up or down, respectively. Since the ketone group is also $sp^2$ hybridized, it adds one $p_z$ electron to the system and can to a first approximation be considered as an additional $p_z$ site on its corresponding sublattice. The atoms at same edges of the chGNRs belong to the same sublattice, whereas the atoms at the opposite edge belong to the other sublattice. A radical pair located at the same/opposite side is therefore expected to be ferromagnetically (FM)/antiferromagnetically (AFM) aligned. All experimentally measured radical pairs match this prediction, except K-S1 (AFM ground state; Table I).

For the conductance spectrum of K-S1, one could argue that the Kondo peak expected from a FM alignment is not visible because its intensity is much lower than that of the spin flip steps (as occurs *e.g.* in K-S0, Fig. 3b) and its width is comparable to J. In fact, the spectrum can be fitted using Ternes' code [50] assuming a FM alignment. However, to do so, the tip-sample transmission function represented by the $T_0^2$ parameter required an anomalously large value. Since we did not change the STM tip during the whole experiments and used comparable tip-sample distances as defined by the STM feedback parameters, it is natural to expect that $T_0^2$ should be similar for all the spin-coupling scenarios. As shown in SM [43], whereas all the other fits rendered comparable $T_0^2$ values, a FM alignment for the K-S1 case required a $T_0^2$ value an order of magnitude higher and clearly out of the error range, suggesting a preferred AFM alignment of the K-S1 radical pair.

This counter-intuitive spin alignment for the K-S1 case is confirmed with DFT calculations, which predict the AFM alignment to be energetically favored. We have additionally performed mean-field Hubbard (MFH) calculations on this same system, given its greater simplicity and successful application to many open-shell carbon nanostructures [23,37]. As discussed earlier, the extra $p_z$ electron on the $sp^2$ hybridized ketone was first considered as an additional $p_z$ site taken to be identical to a carbon $p_z$ orbital. Doing so, MFH calculations with the third nearest-neighbor tight-binding model (3NN-TB) predict a FM ground state for K-S1, in line with the intuitive expectations for radical states on the same ribbon's side but against DFT and experiments [henceforth we call this model MFH-TB(1)]. In a counter-experiment, we performed DFT calculations for another system which is more equivalent to a simulation performed with MFH-TB(1), namely a chGNR with $sp^2$ carbon atoms (*i.e.*, -CH$_2$ groups) instead of the ketones. In this case, the DFT results agree with the intuitive expectations and with the MFH-TB(1), predicting a preferred FM alignment and even a comparable J value [CH$_2$-S1 in Fig. 4(a)]. It follows that the surprising AFM alignment is unequivocally related to the oxygen atom rather than to the mere presence of an extra $p_z$ electron there, exposing insufficient chemical detail in our simple MFH-TB(1) approach.

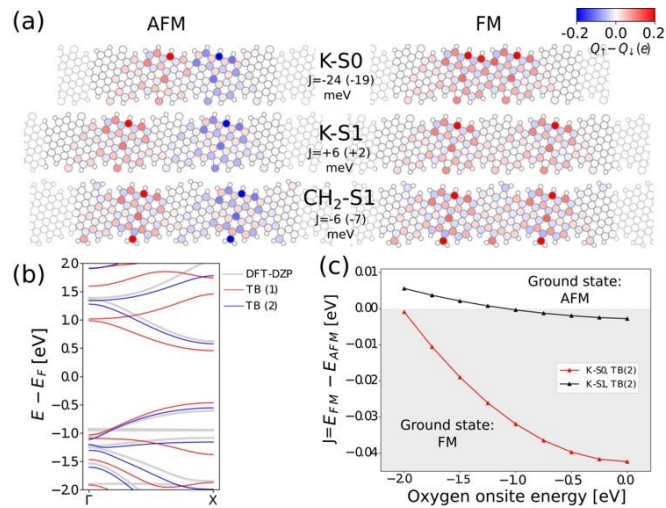

**FIG. 4**. (a) Spin-polarized LDOS distribution of radical states as calculated by DFT and J (MFH-TB(2) values are shown in parentheses) in K-S0, K-S1, and CH$_2$-S1 (all oxygen atoms are replaced by CH$_2$), respectively. (b) Band structures of intact K-chGNR, acquired from DFT, MFH-TB(1), and MFH-TB(2), respectively. (c) MFH-TB calculated J in K-S0 and K-S1 as a function of oxygen onsite energy in TB(2). The onsite Coulomb repulsion in MFH was set to U=3 eV.

In an attempt to make the simple MFH calculations also applicable to these more complicated systems including heteroatoms, we have included modified TB parameters (hopping amplitude and on-site energy) for the oxygen heteroatoms in what we call the model MFH-TB(2). To determine the most appropriate parameters, we compare the calculated band structures obtained from DFT for fully ketone-functionalized chGNRs to those obtained with our MFH-TB models [Fig. 4(b)]. An improvement of the TB model can be obtained with MFH-TB(2) maintaining for carbon atoms the nearest-neighbor hopping amplitude and on-site energies of 2.7 and 0 eV, respectively, but changing the values to 3.8 and -1.5 eV for oxygen atoms bonded to carbon. Qualitatively the increased hopping amplitude on the C=O bonds can be directly related to its shorter length and the non-zero on-site energy to the increased electronegativity of oxygen. The exchange coupling in radical pairs (both strength and sign) is strongly affected by these two parameters, as depicted for the on-site energy in Fig. 4(c) for K-S0 and K-S1. Applying this optimized model now correctly predicts a preferred AFM alignment for K-S1. Furthermore, MFH-TB(2) also corrects the other wrong predictions of the MFH-TB(1) model, as is the open-shell character of the P-O0' and K-O0' radical pairs and provides for all calculated structures J values very similar to DFT calculations (see Table I). Altogether, we hereby provide a simple yet accurate parametrization to model these more complex systems beyond pure hydrocarbon structures.

In summary, we have characterized the magnetic interactions between radical pairs of diverse geometries hosted by two different types of chGNRs, with and without ketone functionalization at their edges. As confirmed by both experiments and theoretical calculations, the exchange coupling between two radicals shows remarkably large variations depending on their relative location on the same or opposite sides of the GNRs, on the spatial distance between them, on the chiral asymmetry, as well as on the structure of the GNR surrounding the magnetic state. These results thus provide valuable information of potential use in the design of graphene-based spin chains and networks with tunable magnetic structures. Furthermore, we demonstrated their modeling by the widely used MFH approximation through a minimal extension to these systems of increased complexity including heteroatoms, which may greatly expedite the understanding and engineering of GNR-based structures in the future.

We thank Dr. Jingcheng Li for fruitful discussions. We acknowledge financial support from the Agencia Estatal de Investigación (grant nos. PID2019-107338RB-C62 and PID2019-107338RB-C63), the European Union's Horizon 2020 research and innovation program (grant no. 863098 and Marie Skłodowska-Curie Actions Individual Fellowship No. 101022150), the Xunta de Galicia (Centro Singular de Investigación de Galicia, 2019-2022, grant no. ED431G2019/03), the European Regional Development Fund, the Spanish Ministry of Economy and Competitiveness (Juan de la Cierva grant No. FJC2019-041202-I), the Basque Government (IT-1255-19), the Basque Departamento de Educación through the PhD scholarship No. PRE_2020_2_0049 (S.S.), the Spanish Research Council (ILINKC20002).

*These authors contributed equally to this work.

†taowang@dipc.org

‡diego.pena@usc.es

§thomas_frederiksen@ehu.eus

‖d_g_oteyza@ehu.eus

Supplemental Material for

# Magnetic Interactions Between Radical Pairs in Chiral Graphene Nanoribbons

Tao Wang,[1,2,*,†] Sofia Sanz,[1,*] Jesús Castro-Esteban,[3] James Lawrence,[1,2] Alejandro Berdonces-Layunta,[1,2] Mohammed S. G. Mohammed,[1,2] Manuel Vilas-Varela,[3] Martina Corso,[1,2] Diego Peña,[3,‡] Thomas Frederiksen,[1,4,§] and Dimas G. de Oteyza[1,2,4,‖]

[1]*Donostia International Physics Center, San Sebastián, Spain*

[2]*Centro de Fisica de Materiales CFM / MPC, CSIC-UPV/EHU, San Sebastián, 20018, Spain*

[3]*Centro Singular de Investigación en Química Biolóxica e Materiais Moleculares (CiQUS) and Departamento de Química Orgánica, Universidade de Santiago de Compostela, Santiago de Compostela, Spain*

[4]*Ikerbasque, Basque Foundation for Science, Bilbao, Spain*

*These authors contributed equally to this work.

†taowang@dipc.org

‡diego.pena@usc.es

§thomas_frederiksen@ehu.eus

‖d_g_oteyza@ehu.eus

Contents

## Methods

**Experiment**

STM measurements were performed using a commercial Scienta-Omicron LT-STM at 4.3 K. The system consists of a preparation chamber with a typical pressure in the low $10^{-10}$ mbar regime and a STM chamber with a pressure in the $10^{-11}$ mbar range. The Au(111) crystal was cleaned *via* two cycles of Ar$^+$ sputtering and annealing (720 K). Pristine 3,1-chGNRs were obtained by deposition of chiral 2,2'-dibromo-9,9'-bianthracene (DBBA) precursor on the Au(111) surface held at room temperature (RT) *via* sublimation at 433 K, followed by a 600 K annealing. K-chGNRs were obtained by deposition of the ketone-DBBA precursors [1] onto Au(111) held at RT *via* sublimation at 448 K, followed by a 670 K annealing. Hydrogenation of the K-chGNR sample was achieved with a hydrogen cracking source with a leak valve. The preparation chamber was first filled to a pressure of $1 \times 10^{-7}$ mbar, after which the tungsten tube was heated to around 2800 K with a heating power of 80 W. The sample was then placed in front of the source for 1 minute and then annealed to 570 K. Oxygenation of P-chGNR sample was achieved by exposing the sample to $3 \times 10^{-5}$ mbar oxygen filled preparation chamber at RT. More detailed information can be found in our previous works [2,3].

All STM and STS measurements shown were performed at 4.3 K. To obtain BR-STM images, the tip was functionalized with a CO molecule that was picked up from the Au(111) surface. CO was deposited onto the sample *via* a leak valve at a pressure of approximately $5 \times 10^{-9}$ mbar and a maximum sample temperature of 7.0 K. CO can be picked up with a metallic tip using a high current and negative bias (*e.g.* I=1 nA, U=−0.5 V). dI/dV measurements were recorded with the internal lock-in of the system. The oscillation frequency used in experiments is 828 Hz. The amplitude for each spectrum is shown in figure captions.

**Theory**

For the DFT with SIESTA [4] calculations we used the DZP basis with an energy shift of 0.02 Ry and the mesh cutoff is of 400 Ry. We used the XC-functional GGA-PBE [5]. We performed calculations with periodic boundary conditions along the horizontal ribbon axis using 2 k-points along the periodic direction. The DFT cell contained 10 precursor units (*i.e.* large extension justifying only 2 k-points). The geometries were determined by relaxing with the expected preferred spin configuration in each case: those geometries with the defects on the same (opposite) side were relaxed with the FM (AFM) configuration. Then, the relaxed geometry is fixed and used to obtain the other spin configuration.

The MFH calculations were performed using the Python code we developed (https://github.com/dipc-cc/hubbard) [6]. We modeled the systems using two different TB models, that we call TB(1) and TB(2), described in the following general Hamiltonian

$$H = \sum_{i\sigma} \varepsilon_{i\sigma} c_{i\sigma}^{\dagger} c_{i\sigma} - \sum_{ij\sigma} t_{ij\sigma} c_{i\sigma}^{\dagger} c_{j\sigma} + U \sum_i (n_{i\uparrow} \langle n_{i\downarrow}\rangle + n_{i\downarrow}\langle n_{i\uparrow}\rangle) \qquad (1)$$

where $c_{i\sigma}$ and $c_{i\sigma}^{\dagger}$ are the creation and annihilation operators for spin index $\sigma = \uparrow, \downarrow$ at site i respectively, and $n_{i\sigma} = c_{i\sigma}^{\dagger} c_{i\sigma}$ is the number operator. The model TB(1) would correspond to simulating the oxygen atoms as CH$_2$: all atoms (including oxygen) are considered carbon atoms (correspondingly saturated with hydrogen atoms) and therefore have the same hopping amplitudes $t_{ij\sigma}$ and onsite energies $\varepsilon_{i\sigma}$. In the model TB(2) we consider the oxygen atoms to have different onsite energy ($\varepsilon^O$=−1.5 eV with respect to the carbon onsite energy which is set to $\varepsilon^C$=0) and the hopping amplitude between C=O first nearest neighboring atoms is of $t_1$ = 3.8 eV, while keeping hopping between C-C atoms $t_1$ = 2.7 eV. In both cases the interaction term is simulated by using U = 3 eV [7,8] to account for the electronic interaction between two $p_z$ orbitals. We consider up to third nearest neighbor hopping interactions, where the hopping between second and third nearest neighbors is ($t_2$, $t_3$=0.2, 0.18) eV [7,8].

# Spin density distributions of radical states on magnetic chGNRs by theoretical calculations

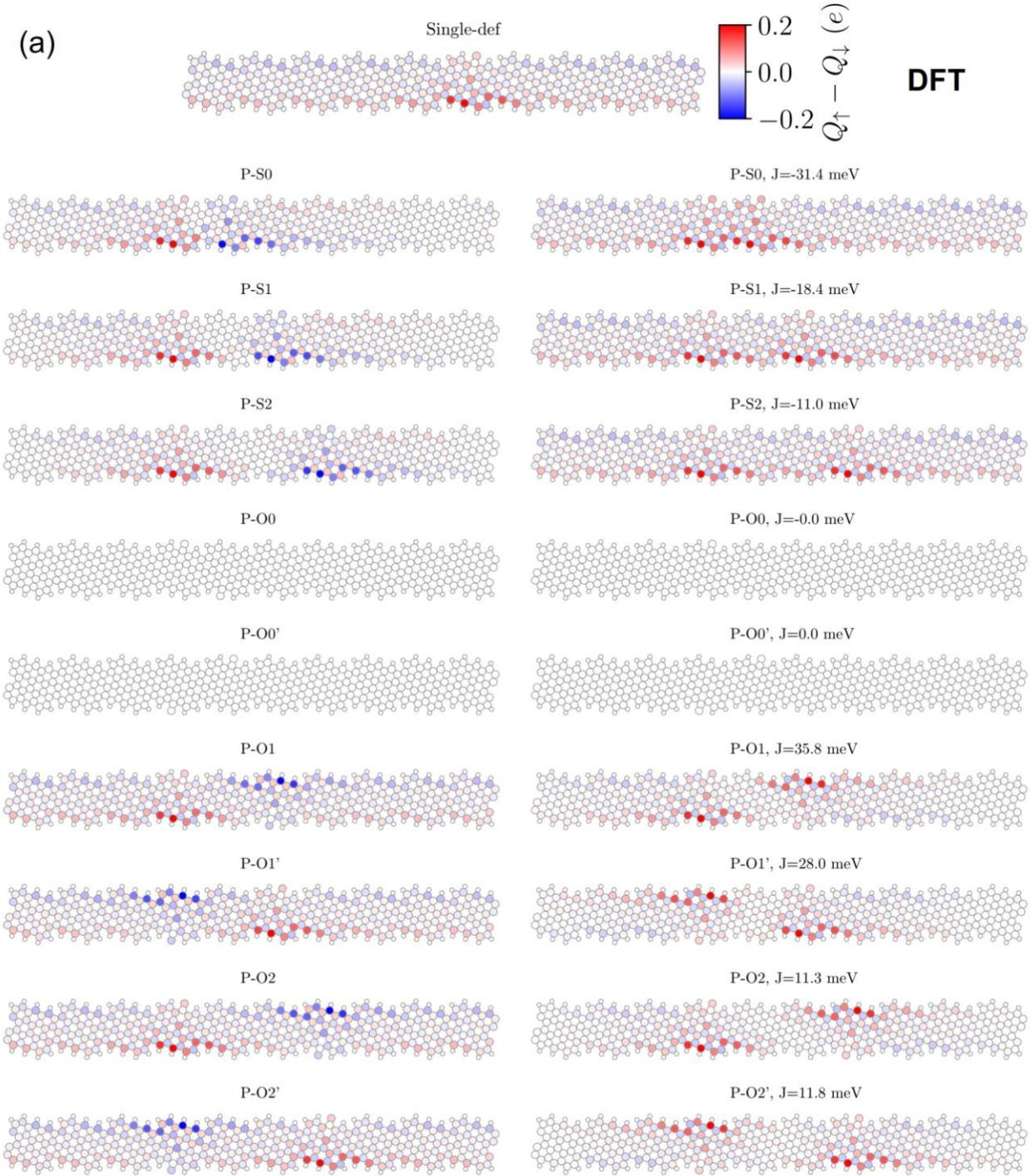

Continued

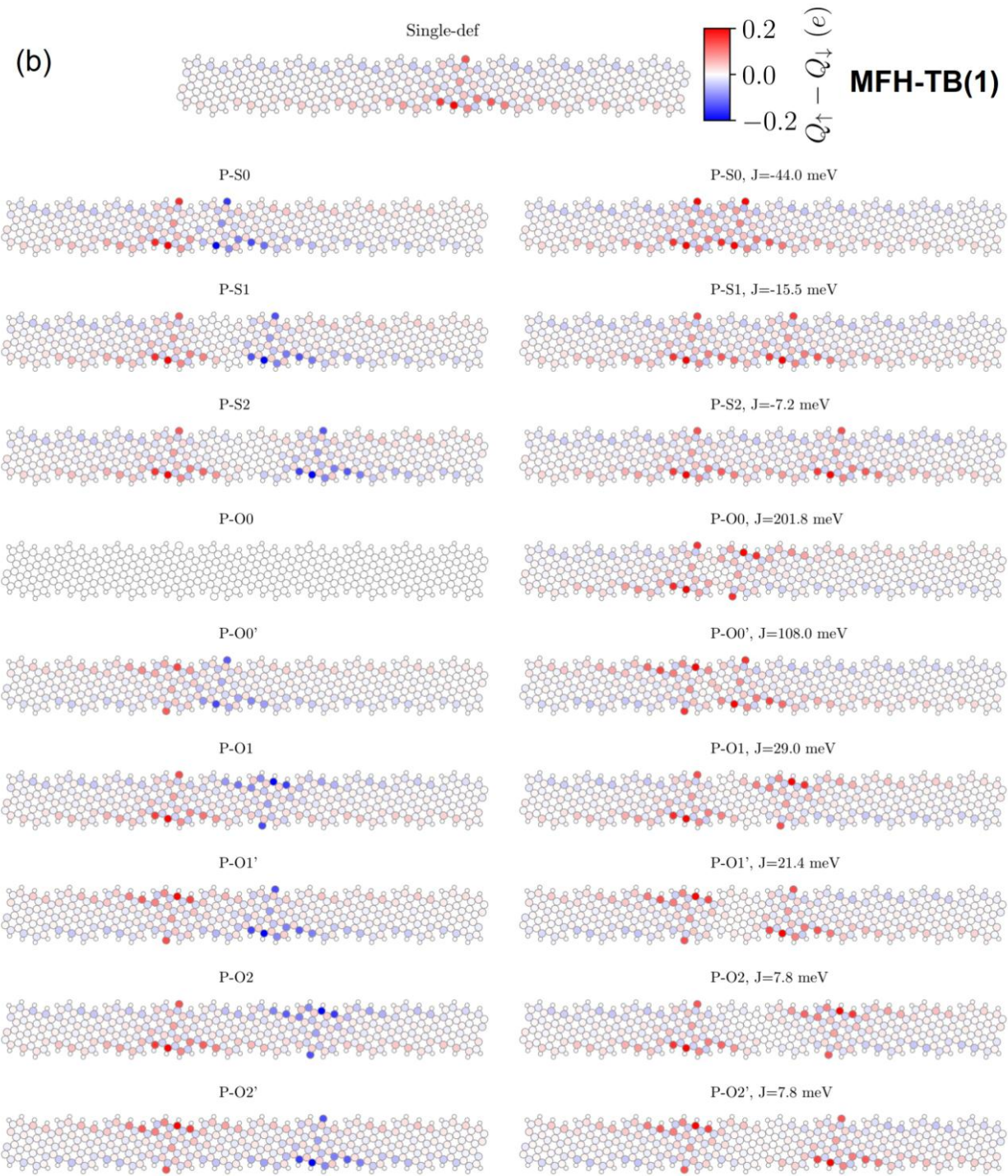

Continued

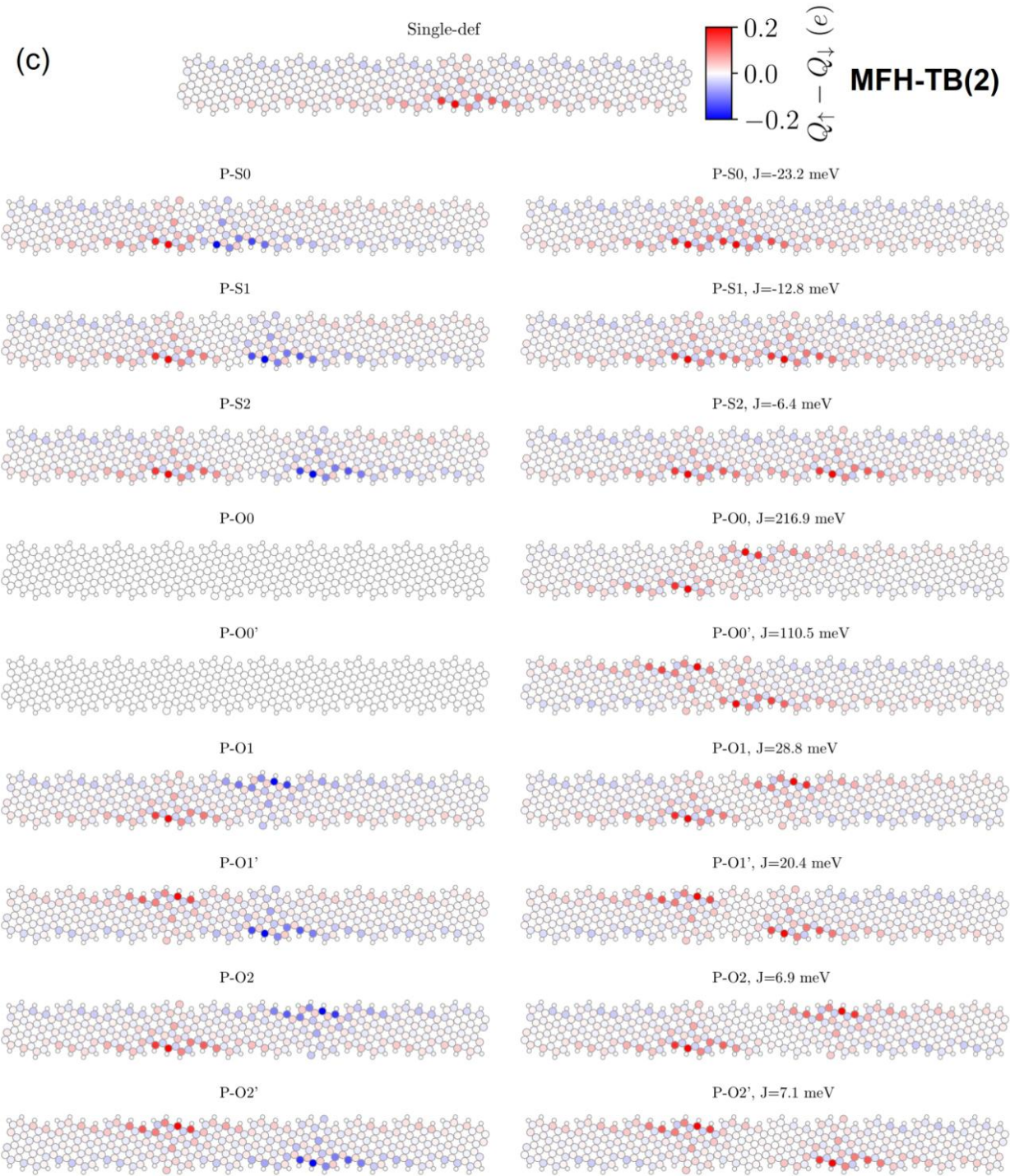

Continued

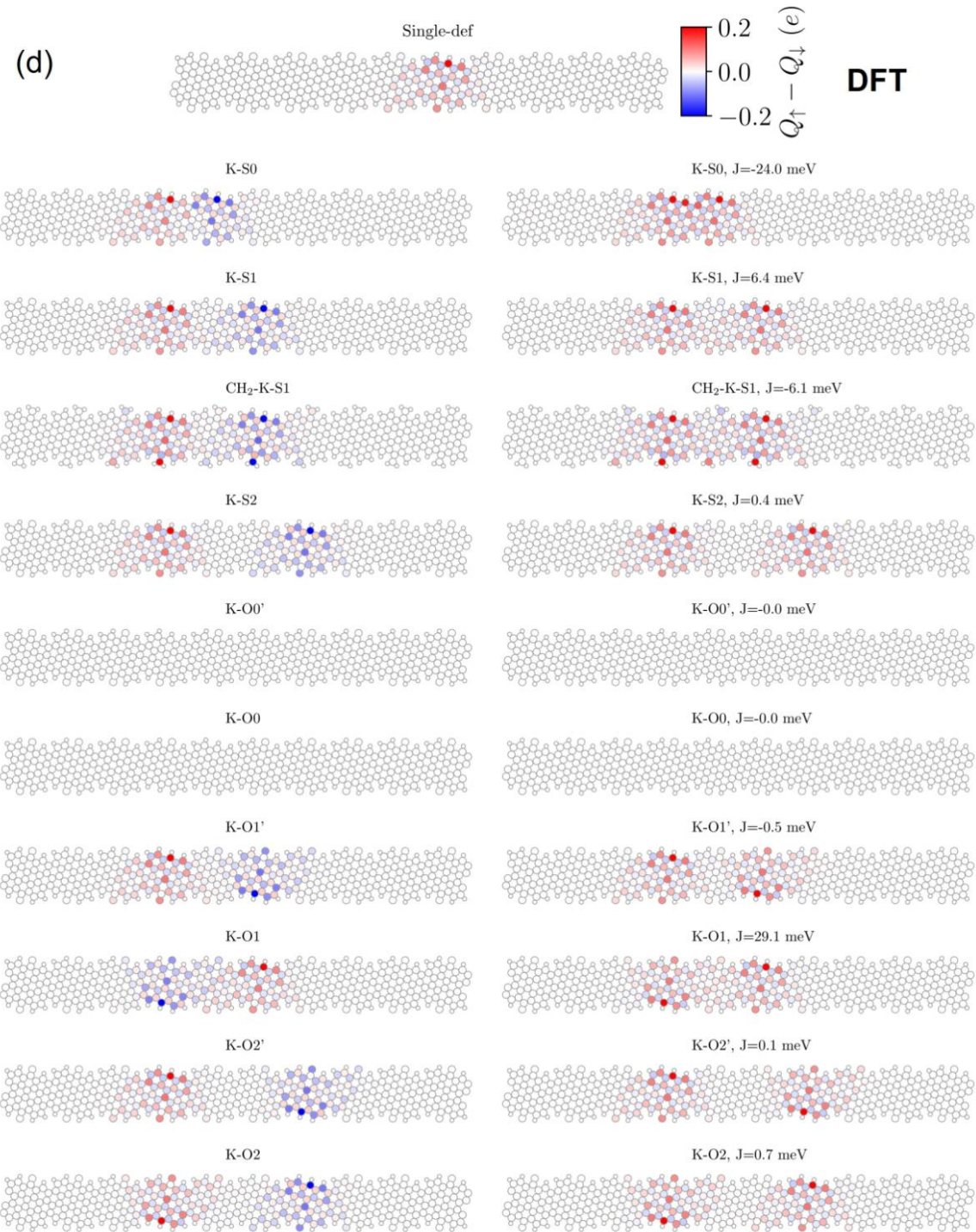

Continued

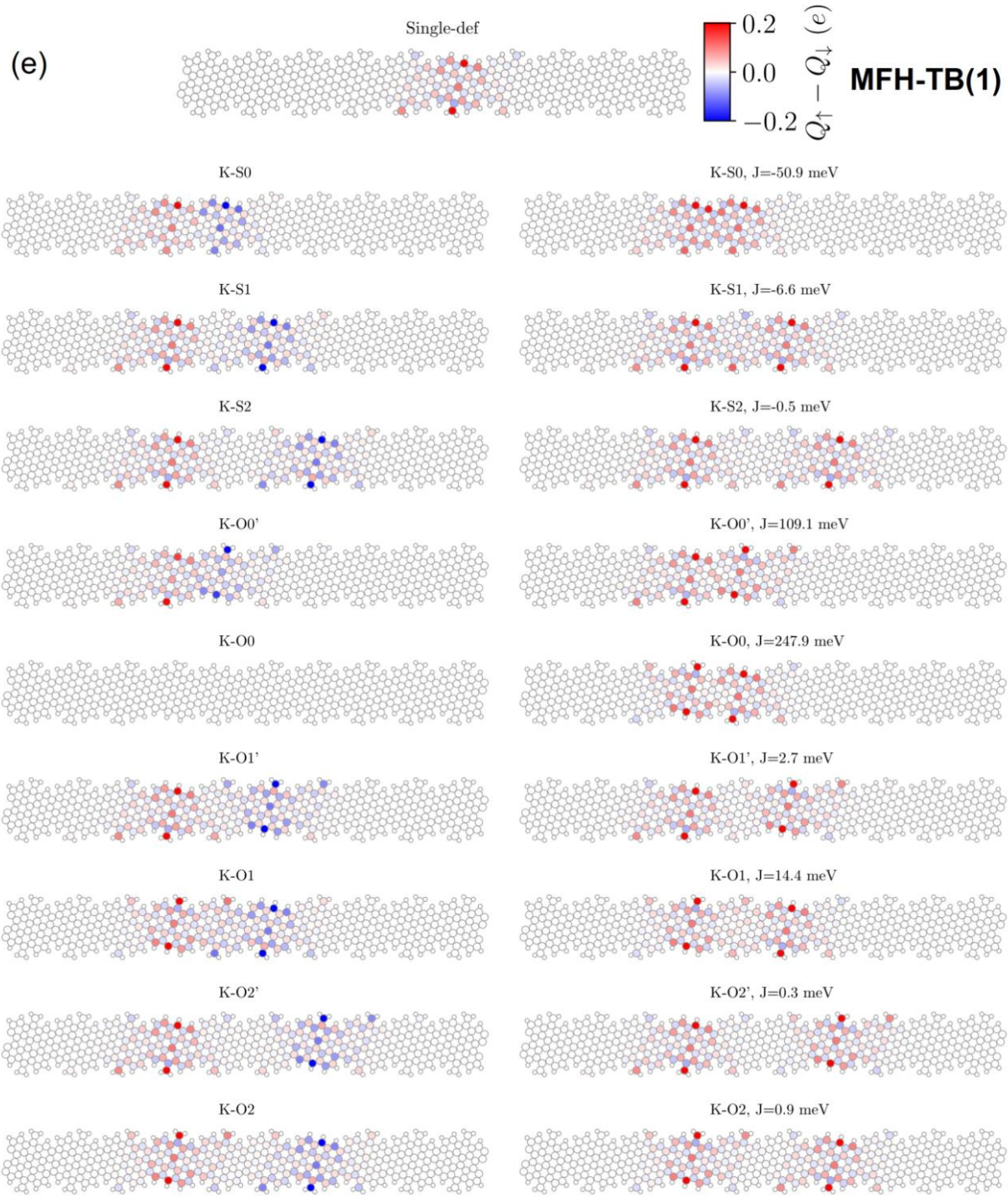

Continued

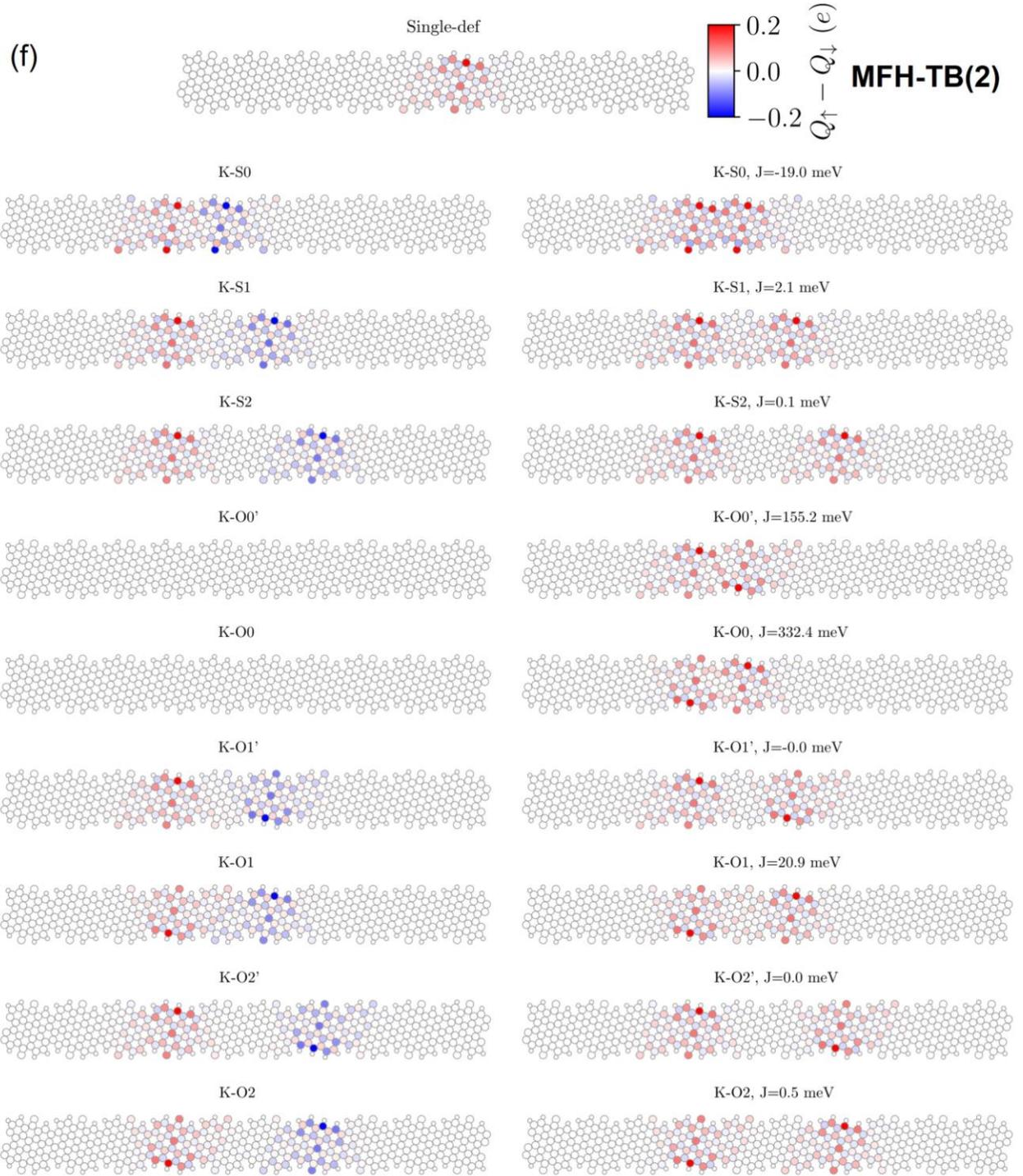

FIG. S1. Spin density distributions of radical states in defective P- and K-chGNRs with a single radical or a radical pair, calculated by DFT, MFH-TB(1), and MFH-TB(2), respectively.

**Characterization of the magnetism of a P-chGNR with a single defect**

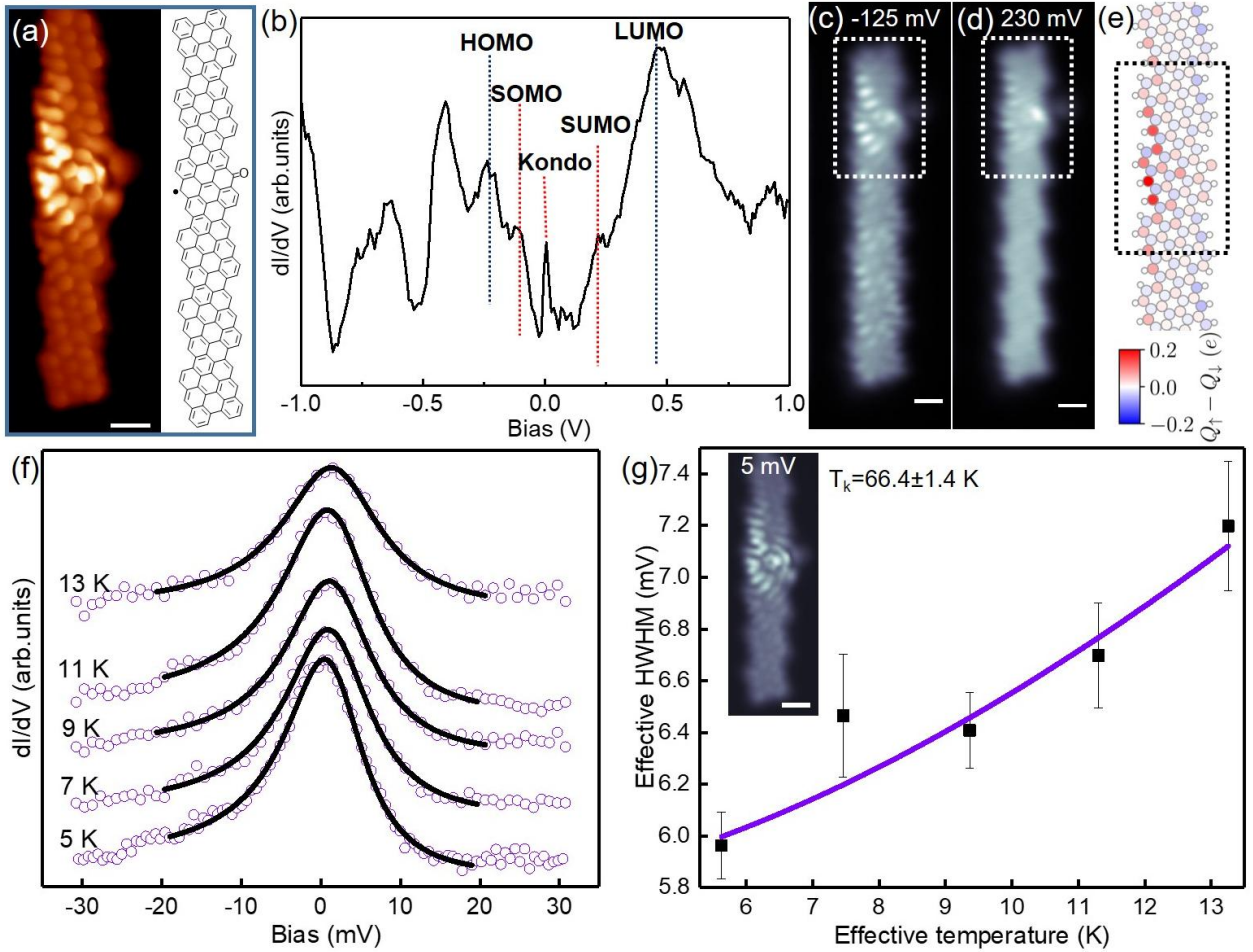

FIG. S2. (a) Constant-height BR-STM image of P-chGNR with a single magnetic defect, acquired by a CO-functionalized probe, together with the corresponding chemical structure. Bias: 5 mV. (b) dI/dV spectra taken at the marked position in (a), $V_{rms}$= 20 mV. (c,d) Constant-height dI/dV maps taken with a CO functionalized tip, recorded at the bias of −0.47 and 0.25 V, respectively. $V_{rms}$= 20 mV. The energy of SOMO is close to HOMO thus the signal from HOMO is also visible in (c). (e) DFT calculated spin density distribution in P-chGNR with a single magnetic defect. The dashed frames in (c-e) include same units with high spin density distributions. (f) Temperature dependence of the Kondo resonance between the unpaired π radical and conduction electrons on the Au(111) substrate. All the spectra at different temperatures are fitted by a Frota function. (g) Extracted half-width at half-maximum (HWHM) of the Kondo resonances as a function of temperature, fitted by the Fermi-liquid model: HWHM=$\frac{1}{2}\sqrt{(\alpha k_B T)^2 + (2k_B T_K)^2}$. The HWHM is corrected to the effective HWHM considering the broadening contributed from the tip temperature to the intrinsic HWHM. The temperature of the sample is also corrected to the effective temperature considering the lock-in oscillation amplitude contribution [9]. A Kondo temperature of 66.4 K is obtained. The Kondo temperature is similar to that in K-chGNR with a single defect (66.0 K). A Kondo current map of a P-chGNR with a single magnetic defect is inserted, which was acquired at 5 mV in constant-height mode using relatively far tip-sample distance. All the scale bars in STM images are 0.5 nm.

## Closed-shell structures

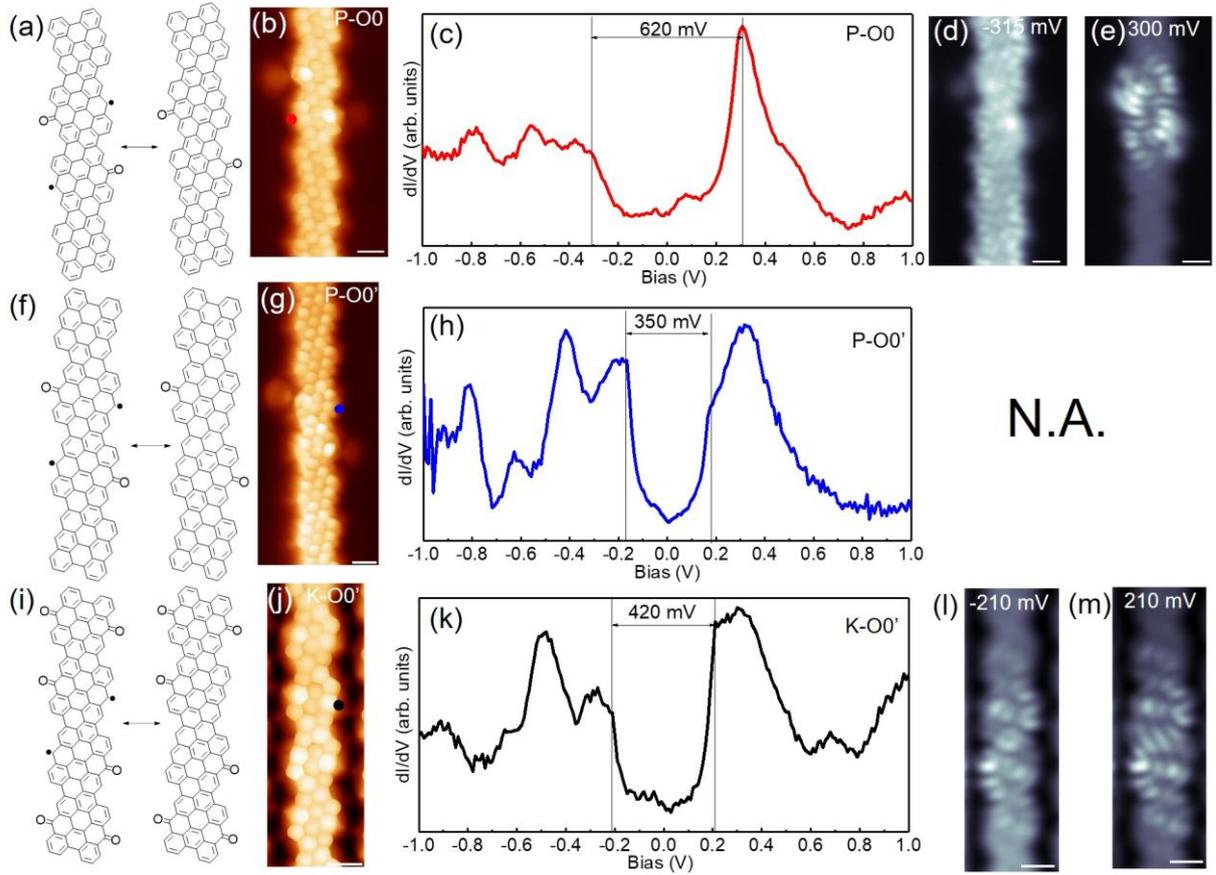

FIG. S3. (a,f,i) Open-shell and closed-shell resonant structures of P-O0, P-O0', and K-O0'. (b,g,j) Constant-height STM images of P-O0, P-O0', and K-O0' GNRs acquired using CO functionalized probes. V=5 mV. All the scale bars are 0.5 nm. (c,h,k) dI/dV spectra taken at the marked positions in (b,g,j), respectively. (d,e) dI/dV maps acquired at −315 mV and 300 mV for P-O0, respectively. (l,m) dI/dV maps acquired at −210 mV and 210 mV for P-O0, respectively. Lock-in amplitude is 20 mV for all the dI/dV maps. Note that SOMO and SUMO separated by the spin splitting with a Coulomb gap show identical LDOS. The different distributions of occupied and unoccupied orbitals demonstrate they are doubly occupied HOMOs and LUMOs.

**Kondo peak comparison between K-chGNR with single defect and K-S2, K-O2'**

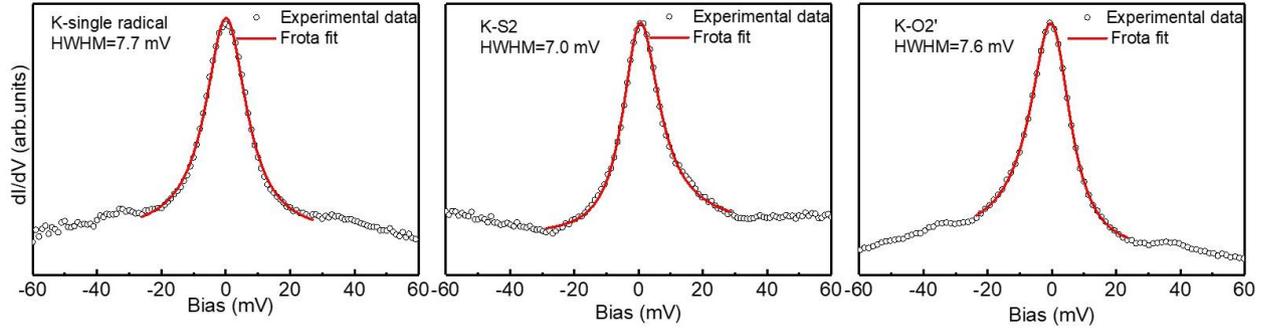

FIG. S4. Comparison between Kondo peaks of K-chGNR with a single radical, K-S2 and K-O2'. Similar HWHMs are obtained by fitting the three peaks with a Frota function. All the spectra were acquired with CO functionalized probes.

**Simulations of the dI/dV spectra from radical pairs**

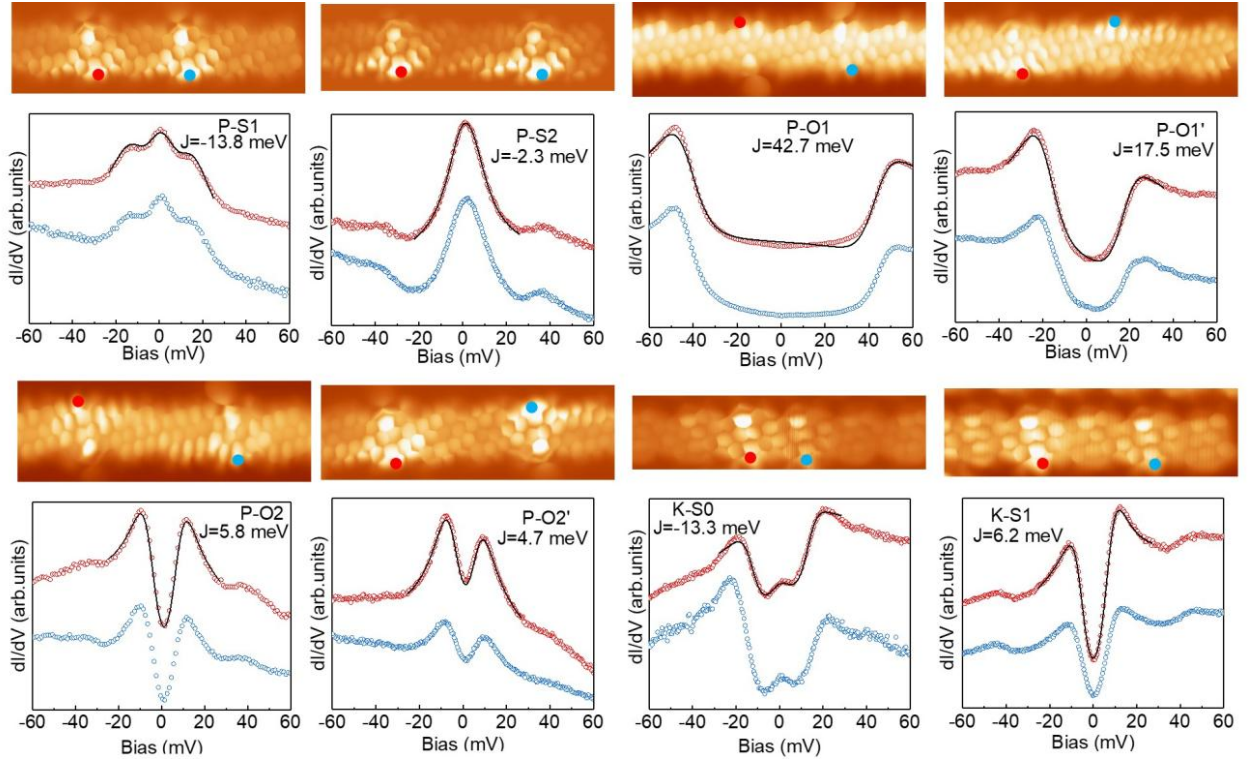

FIG. S5. Comparison between simulated (black) and measured (red and blue) spectra taken for coupled radicals in defective chGNRs. We have chosen the red spectra for the simulation using the code from M. Ternes [10]. The simulated spectra (black lines) are overlaid on the red spectra. In each case the J values obtained from the fitting is written alongside with the radical pair type. All the scale bars in the STM images are 0.5 nm.

**Simulation of K-S1 by a FM ground state**

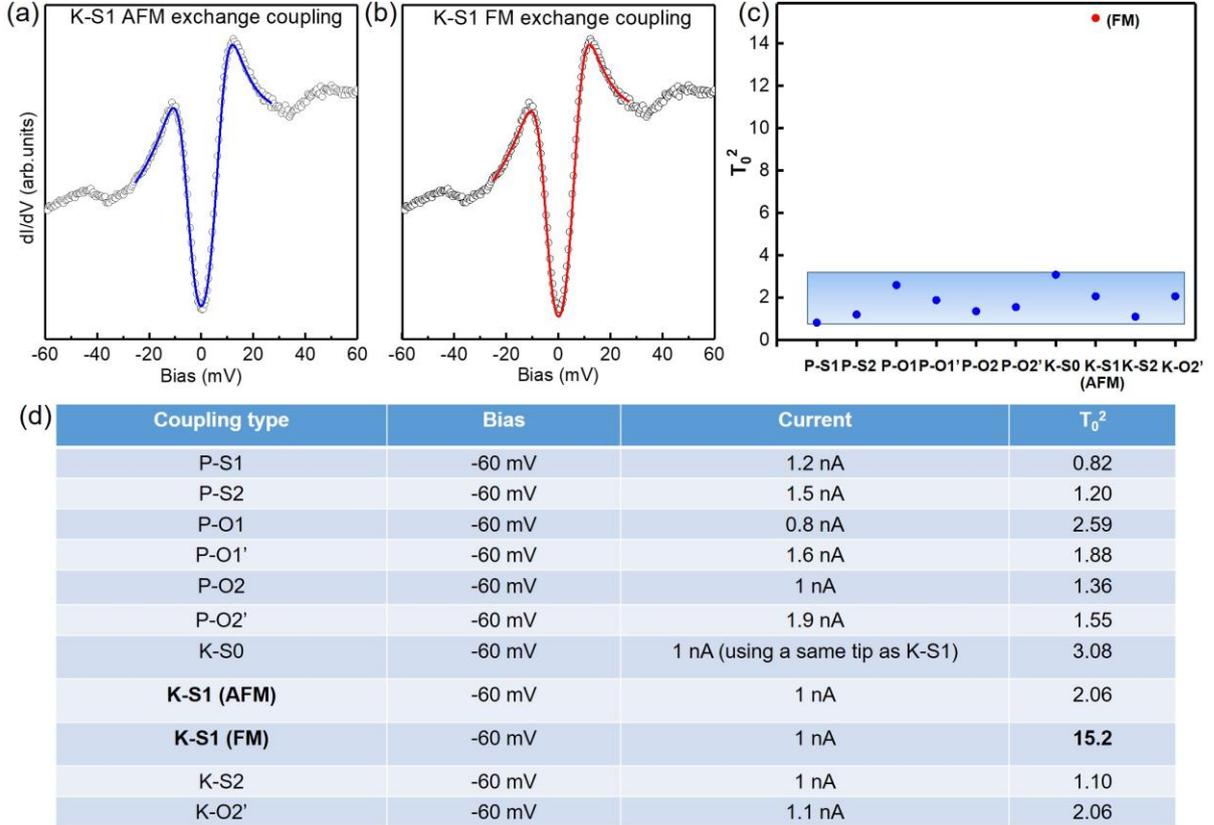

FIG. S6. (a,b) Simulations of K-S1 by AFM and FM ground states respectively, using the code from Ternes [10]. (c) Tunnel barrier transmission coefficient $T_0^2$ values of all the spin coupling cases observed in experiments, including both FM and AFM coupling of K-S1, acquired from the simulations in (a,b) and FIG. S5. The $T_0^2$ value FM KS1 obviously exceeds its reasonable range (the blue frame) as obtained from other cases. (d) The correlated bias and current values used for dI/dV measurement as well as $T_0^2$ values for each case. Since the currents for all the cases are similar, the obtained $T_0^2$ values should be also similar. However, the $T_0^2$ value from FM coupled K-S1 is abnormally high, thus the possibility can be excluded.